\documentclass[pra,twocolumn,floatfix,superscriptaddress,longbibliography, nofootinbib]{revtex4-1}
\usepackage{amsmath,amsfonts,amssymb,amsthm,graphics,graphicx,epsfig,bbm}
\usepackage[colorlinks=true,citecolor=blue,linkcolor=blue,urlcolor=blue]{hyperref}
\usepackage[usenames]{color}
\usepackage[dvipsnames]{xcolor}
\usepackage{graphicx}
\usepackage{tikz}
\usepackage{subfigure}
\usepackage{float}
\usepackage{amsmath}
\usepackage{epsfig}
\usepackage{dcolumn}
\usepackage{bm}
\usepackage{color}
\usepackage{times}
\usepackage{epstopdf}
\usepackage{amscd}
\usepackage{amsthm}
\usepackage{amssymb}
\usepackage{bbm}
\usepackage{amssymb}
\usepackage{amstext}
\usepackage{latexsym}
\usepackage{comment}
\usepackage{hyperref}
\usepackage{amsfonts}
\usepackage{psfrag}
\usepackage{soul,xcolor}
\usepackage[normalem]{ulem}
\usepackage{dsfont}
\usepackage{txfonts}
\usepackage{footnote}
\usepackage{multirow}
\usepackage{appendix}
\usepackage{mathtools}
\usepackage{tikz}
\usetikzlibrary{quantikz}%to draw quantum circuits
\usepackage{bbold}
\usepackage{enumitem}
\usepackage{xstring}
\usepackage{cancel}
\usepackage{amscd}
\usepackage{bbm}
\usepackage{units}
\usepackage{cancel}
\usepackage{xstring}
\usepackage{import}
\usepackage{subfiles}
\usepackage{amsmath,amsfonts,amssymb,amsthm,graphics,graphicx,epsfig,bbm}
\usepackage[colorlinks=true,citecolor=blue,linkcolor=blue,urlcolor=blue]{hyperref}
\usepackage[usenames]{color}
\usepackage{graphicx}
\usepackage{subfigure}
\usepackage{amsmath}
\usepackage{epsfig}
\usepackage{dcolumn}
\usepackage{bm}
\usepackage{color}
\usepackage{times}
\usepackage{epstopdf}
\usepackage{amssymb}
\usepackage{amstext}
\usepackage{latexsym}
\usepackage{hyperref}
\usepackage{amsfonts}
\usepackage{psfrag}
\usepackage{soul,xcolor}
\usepackage[normalem]{ulem}
\usepackage{dsfont}
\usepackage{txfonts}
\usepackage{xcolor}
\usepackage{footnote}
\usepackage{multirow}
\usepackage{appendix}
\usepackage{mathtools}
\usepackage{ulem}
\usepackage{mathtools}

\definecolor{mycolor}{RGB}{106,81,162}

\usepackage{mathtools}

\newcommand{\iden}[1]{
    \ifthenelse{\equal{1}{\string #1}}
  {% True case
   \mathbbm{1}
  }
  {% false case
   \mathbbm{1}^{\otimes#1}}
  }
\newcommand{\ketzero}[1]{
    \ifthenelse{\equal{1}{\string #1}}
  {% True case
   \ket{0}
  }
  {% false case
   \ket{0}^{\otimes#1}}
  }
\newcommand{\brazero}[1]{
    \ifthenelse{\equal{1}{\string #1}}
  {% True case
   \bra{0}
  }
  {% false case
   \bra{0}^{\otimes#1}}
  }
\newcommand{\ketone}[1]{
      \ifthenelse{\equal{1}{\string #1}}
    {% True case
     \ket{1}
    }
    {% false case
     \ket{1}^{\otimes#1}}
    }
  \newcommand{\braone}[1]{
      \ifthenelse{\equal{1}{\string #1}}
    {% True case
     \bra{1}
    }
    {% false case
     \bra{1}^{\otimes#1}}
    }

\usepackage{todonotes}

\usepackage{dsfont}
\usepackage{amssymb}
\usepackage{txfonts,color}
\usepackage{multirow}
\usepackage{booktabs} % For prettier tables
\makeatother

\begin{document}

\title{Satellite image classification with neural quantum kernels}

\author{Pablo Rodriguez-Grasa}
\email[Corresponding author: ]{\qquad pablojesus.rodriguez@ehu.eus}
\affiliation{Department of Physical Chemistry, University of the Basque Country UPV/EHU, Apartado 644, 48080 Bilbao, Spain}
\affiliation{EHU Quantum Center, University of the Basque Country UPV/EHU, Apartado 644, 48080 Bilbao, Spain}
\affiliation{TECNALIA, Basque Research and Technology Alliance (BRTA), 48160 Derio, Spain}

\author{Robert Farzan-Rodriguez}
\affiliation{Department of Artificial Intelligence and Big Data, GMV, Isaac Newton 11, Tres Cantos, 28760 Madrid, Spain}

\author{Gabriele Novelli}
\affiliation{Department of Artificial Intelligence and Big Data, GMV, Isaac Newton 11, Tres Cantos, 28760 Madrid, Spain}

\author{Yue Ban}
\affiliation{Departamento de F\'isica, Universidad Carlos III de Madrid, Avda. de la Universidad 30, 28911 Legan\'es, Spain} 
\affiliation{Instituto de Ciencia de Materiales de Madrid (CSIC), Cantoblanco, E-28049 Madrid, Spain}
\affiliation{TECNALIA, Basque Research and Technology Alliance (BRTA), 48160 Derio, Spain}

\author{Mikel Sanz}
\affiliation{Department of Physical Chemistry, University of the Basque Country UPV/EHU, Apartado 644, 48080 Bilbao, Spain}
\affiliation{EHU Quantum Center, University of the Basque Country UPV/EHU, Apartado 644, 48080 Bilbao, Spain}
\affiliation{IKERBASQUE, Basque Foundation for Science, Plaza Euskadi 5, 48009, Bilbao, Spain}
\affiliation{Basque Center for Applied Mathematics (BCAM), Alameda de Mazarredo, 14, 48009 Bilbao, Spain}

\begin{abstract}

    Achieving practical applications of quantum machine learning for real-world scenarios remains challenging despite significant theoretical progress. This paper proposes a novel approach for classifying satellite images, a task of particular relevance to the earth observation (EO) industry, using quantum machine learning techniques. Specifically, we focus on classifying images that contain solar panels, addressing a complex real-world classification problem. Our approach begins with classical pre-processing to reduce the dimensionality of the satellite image dataset. We then apply neural quantum kernels (NQKs)—quantum kernels derived from trained quantum neural networks (QNNs)—for classification. We evaluate several strategies within this framework, demonstrating results that are competitive with the best classical methods. Key findings include the robustness of or results and their scalability, with successful performance achieved up to 8 qubits.

\end{abstract}
\maketitle

\section{Introduction}
Recent advancements in the field of quantum machine learning (QML) \cite{QML, qml_physical_sciences, qml_vedran} have significantly enhanced our understanding of how quantum resources can be utilized to design various new paradigms \cite{power_of_data, schuld2021supervised,qmlbkm, qml_in_feature}. These developments have shed light on the trainability \cite{bp_mcclean, bp_cost_function, bp_ansatz_expressibility, bp_adjoint, bp_unified, kernel_concentration, absence_QCNN} and generalization capabilities of quantum models \cite{generalization_encoding, generalization_few_training, generalization_overfitting, generalization_QML, understanding_QML} and have even demonstrated theoretical advantages over classical methods for certain tailored problems \cite{advantage_establishing, advantage_qvsc, advantage_reinforcement, advantage_rigorous, advantage_superpoly}. Despite these breakthroughs and the successful implementation of quantum models on quantum processors \cite{imple_quantum_enhanced, implement_nearest_centroid, implementation_advantage, implementation_QONN,peters2021machine}, the practicality of QML remains uncertain \cite{real_world}.

Image classification is one of the most usfeul tasks due to its significant industrial applications. Various approaches have been explored, including quantum convolutional neural networks (QCNNs)\cite{qcnn_quantum_classical, quanvolutional, a_qcnn_for_image, qcnn_for_image, simen2024digitalanalogquantumconvolutionalneural, altares_QCNN} and models that leverage symmetries \cite{sym_approximately, sym_enh_im, sym_paul, sym_provably}. There is increasing recognition that combining classical and quantum approaches can be more beneficial than trying to replace one by the other. The pursuit of better machine learning techniques with new methodologies involves a deeper consideration of various factors such as resource efficiency, data requirements, and the practical applications of these technologies. 
As addressed in this work, this involves pre-processing images classically before applying a quantum data processing model \cite{qml_for_image, k_NN}. The aim is to demonstrate the development and implementation of models using components executable on quantum processors, with the goal of scaling them up to create models that cannot be simulated on classical processors.

Most studies in this context focus on typical classification problems that serve as benchmarks for evaluating these models, while just a few address more realistic classification challenges. 
Earth observation (EO), with the use of remote sensing technologies, provides valuable data for a wide range of applications, including climate change monitoring, disaster management, agricultural productivity assessment, and urban planning. By using machine learning, EO has become essential for extracting useful insights from this data, helping us manage Earth's systems more effectively. Recently, the daily volume of satellite image products has surged significantly due to a wide range of institutional and commercial missions \cite{eoportal, copernicus_data}. This growth has prompted the EO community to explore computational paradigms that could enhance current methods.

Few studies have addressed complex classification tasks using quantum resources; this work aims to bridge that gap by combining classical and quantum resources to demonstrate that quantum methods can achieve effective results in real-world scenarios. Our focus is on the identification of solar photovoltaic (PV) panels, utilizing a challenging dataset from a real-world scenario \cite{distributed-dataset} to address a high-priority social and environmental application. In order to achieve this, we developed an effective classical pipeline tailored to preprocess images from this complex satellite imagery dataset, reducing them to a limited number of features suitable for encoding in a quantum architecture. Building on previous studies that underscore the dependency of results on ansatz selection in approaches using parameterized quantum circuits \cite{image_quantum, circuit_based, benítezbuenache2024bayesian, pqgates} or on the choice of embedding in support vector machines based on quantum kernels \cite{MOUNTRAKIS2011247, 9554802, qsvm_ZZ, miroszewski2023quantum, altares_kernel}, we chose to employ neural quantum kernels (NQKs) \cite{rodriguezgrasa2024training}. This model uses a quantum neural network to generate a embedding quantum kernel (EQK) tailored to the classification task.

The decision to utilize NQKs addresses several challenges in the current quantum machine learning landscape. Firstly, it introduces a scalable model that mitigates trainability issues known in parametrized quantum circuits and quantum neural networks \cite{McClean_2018, Cerezo_2021, Holmes_2022, Anschuetz_2022, ragone2023unified}. As we show in this work, we achieve consistently improved results with the addition of qubits. Additionally, since training quantum neural networks can be problematic, constructing a kernel from the network has proven to be resilient to these training difficulties. This resilience allows us to demonstrate that constructing an effective kernel model does not require the ability to finely tune a quantum neural network. Finally, while the impact of entanglement on enhancing quantum machine learning models remained uncertain \cite{bowles2024betterclassicalsubtleart}, our findings demonstrate that the performance improvements observed when adding qubits to neural quantum kernels are solely due to the presence of entanglement. This clarifies the crucial role of entanglement in these models.

Within the framework of NQKs, we consider two model constructions. The first, the $1$-to-$n$ model, is built by training a single-qubit QNN and then using the trained parameters to construct an $n$-qubit EQK. With this model, we observe that using just 2–3 qubits yields results comparable to classical counterparts, and we show its robustness with respect to the training of the single-qubit QNN. The second model, the $n$-to-$n$ model, involves training an $n$-qubit QNN by iteratively adding qubits to ensure scalability, then using the trained architecture to construct the corresponding $n$-qubit EQK. In this case, the trained QNN functions directly as the quantum embedding. We show that results improve progressively as qubits are added. With both models, we achieve accuracies approaching 90\%.

This paper is organized as follows: Section~\ref{sec:dataset} outlines the dataset generation process, explaining how \( p \) features are extracted from the original images to serve as inputs for the quantum model. In Section~\ref{sec:methodology}, we introduce Neural Quantum Kernels (NQKs), a quantum classification approach proposed in Ref.~\cite{rodriguezgrasa2024training}. Section~\ref{sec:results} presents and analyzes the numerical results. Section~\ref{discussion} explores the implications and limitations of our work, proposing directions for future research. Finally, Section~\ref{sec:conclusion} summarizes our findings, providing a concise conclusion to the manuscript.

\section{Related Work}
\label{sec:related_work}

QML techniques have shown potential in advancing remote sensing image classification by addressing challenges like high-dimensional data and leveraging quantum computational power. This section provides an integrated overview of key contributions in feature extraction and dimensionality reduction, hybrid QNNs, and  support vector machines based on quantum kernels.

Feature extraction and dimensionality reduction play a critical role in adapting high-resolution satellite images to the limited qubit capacity of NISQ computers. Studies such as~\cite{pqgates} and~\cite{image_quantum} have demonstrated the effectiveness of classical preprocessing methods like convolutional autoencoders (CAE) and principal component analysis (PCA) for reducing dimensionality before quantum processing. Notably, CAE-based techniques were found to outperform naive downsampling methods, embedding relevant features into parameterized quantum circuits. Furthermore,~\cite{henderson2020methods} introduced quantum convolutional filters trained through variational circuits, demonstrating their efficacy in classifying datasets like DeepSat SAT-4 using real quantum hardware.

Hybrid QNNs, combining classical neural networks with quantum circuits, have emerged as a promising approach for Earth observation. Works such as~\cite{circuit_based} and~\cite{9553133} utilized PQCs within classical convolutional pipelines to encode and classify RGB EuroSAT and Onera Satellite Change Detection datasets. These studies highlighted the importance of convolutional embeddings and explored various ansatz, with real amplitude circuits offering compact architectures but slightly lower performance compared to ResNet-based solutions. Additionally,~\cite{transfer} explored quantum transfer learning by freezing classical layers and training quantum layers, finding that strongly entangled multi-qubit circuits performed best. Hyperparameter tuning in hybrid QNNs, a complex yet critical aspect, was systematically addressed in~\cite{hyperparameters}, providing strategies for optimizing circuit depth and qubit count using EuroSAT as a benchmark.

Various studies have explored the use of quantum kernels for solving real-world tasks, though these kernels are typically problem-agnostic \cite{Huang_2021, peters2021machine, Bartkiewicz_2020, Kusumoto_2021, Wu2023quantumphase, kyriienko2022unsupervisedquantummachinelearning}. The development of neural quantum kernels, however, opens the door to creating quantum kernels specifically tailored to individual problems. In the context of satellite image classification, both gate-based circuits and quantum annealing have been applied, as demonstrated in~\cite{qsvm_ZZ}, where these methods achieved performance comparable to classical SVMs on multispectral data. Despite this, real hardware limitations imposed constraints on data scalability. Recent advancements, such as those in~\cite{miroszewski2023quantum}, revealed that while both classical and quantum runtimes scale logarithmically with dimensionality, real quantum hardware can offer runtime advantages over simulations, suggesting potential for intermediate-scale quantum systems.

Collectively, these works underline the synergistic potential of combining classical preprocessing with quantum methodologies, while also highlighting the challenges and opportunities in developing scalable and efficient QML frameworks for remote sensing applications.

%% SECTION: THE DATASET %%

\section{The Dataset}
\label{sec:dataset}

This study utilizes an annotated dataset comprising of over 19,000 solar panels distributed across 601 high-resolution satellite images taken from four cities in California. The dataset was originally introduced in Ref.~\cite{distributed-dataset}. The dataset has been extensively used in the fields of machine learning and deep learning for tasks such as solar forecasting (estimating solar power generation), environmental and socioeconomic analyses, as well as mapping and detection of PV installations~\cite{ERDENER2022112224, FENG2021107176, YU20182605}.

\begin{figure}[t!]
    \centering
    \includegraphics[width=0.85\columnwidth]{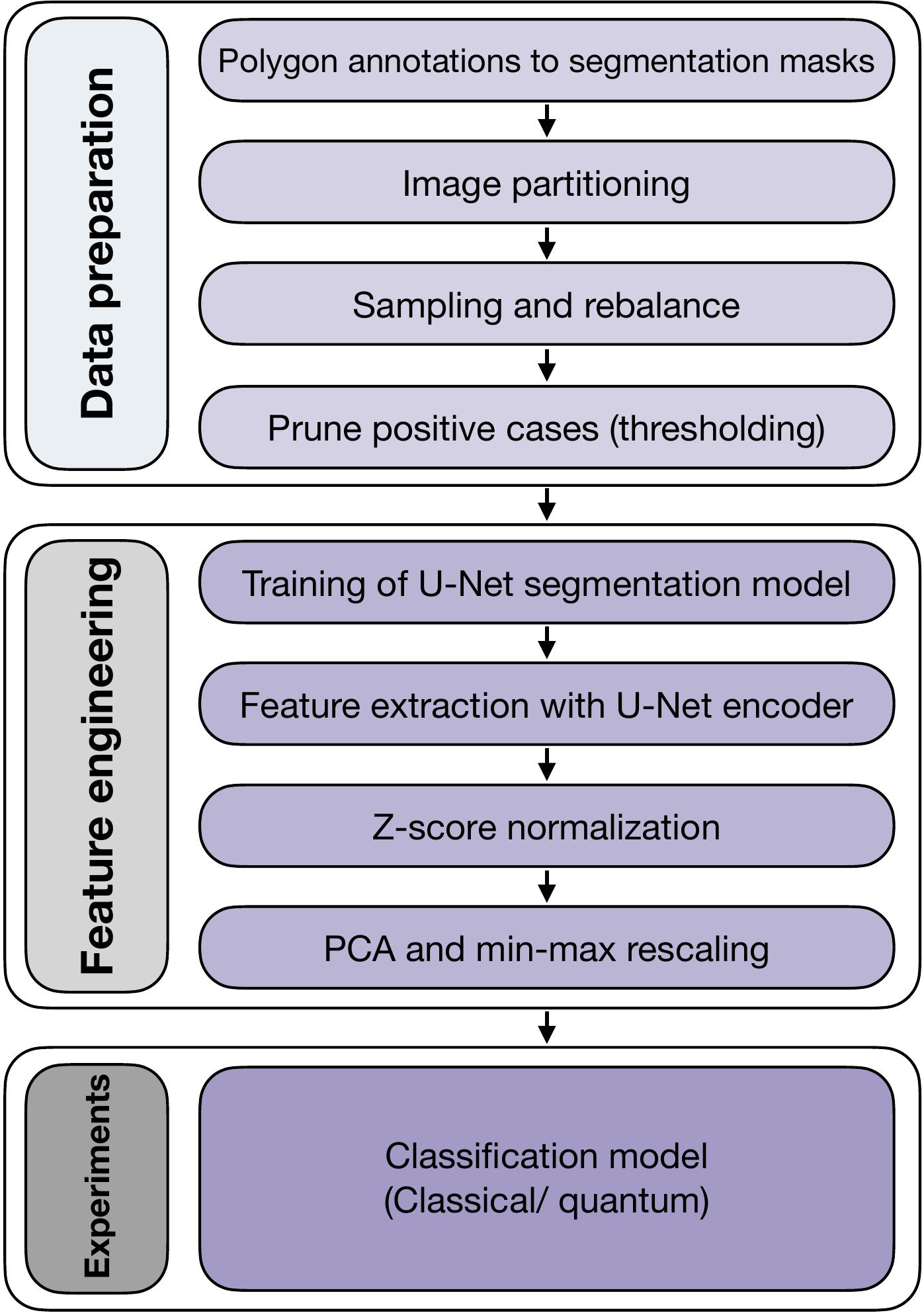}
    \caption{Flowchart of the image processing workflow. The input dataset for the classification model is prepared, and features are extracted to meet the requirements for encoding the information into a quantum model.}
    \label{fig:pipeline-pca}
\end{figure}

The following subsections, \ref{sub:data_prep} and \ref{sub:feat_eng}, report details about the dataset preparation and feature engineering steps, required for the following implementation of classical and quantum classification models. The whole workflow is shown in Figure \ref{fig:pipeline-pca}.

\subsection{Data preparation}\label{sub:data_prep}

%----------------------FIGURE 1-------------------------
%-------------------------------------------------------
\begin{figure}[t!]
    \centering
    \includegraphics[width=0.85\columnwidth]{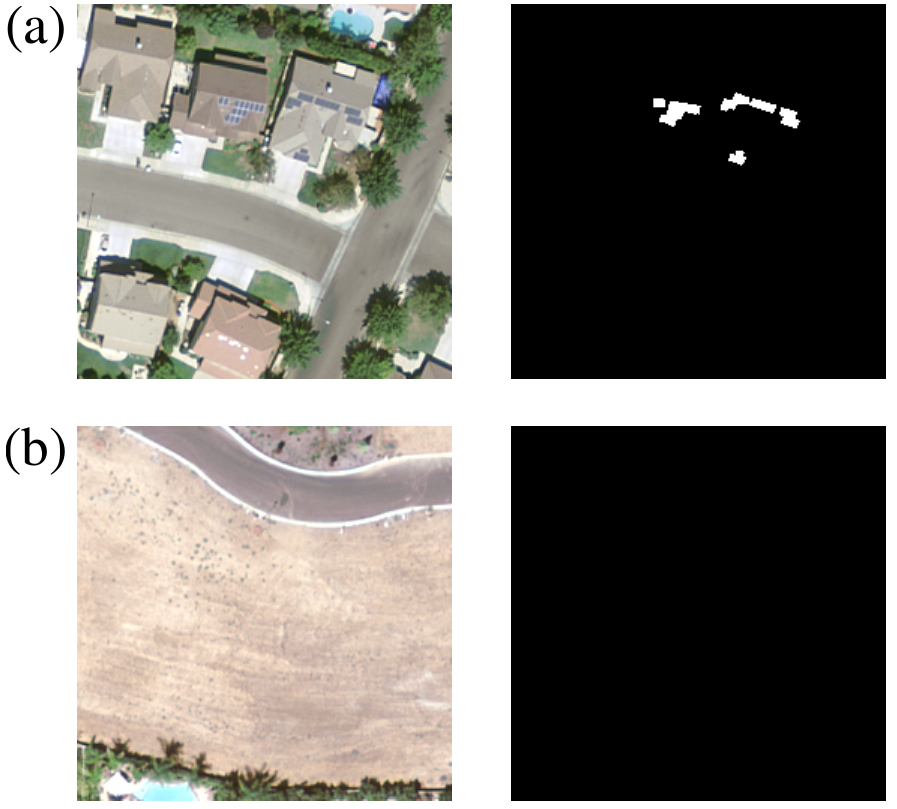}
    \caption{(a) Positive image sample with solar panels ($\hat{y} = +1$) and (b) negative image sample without solar panels ($\hat{y} = -1$), each accompanied by their respective binary segmentation masks. The negative sample illustrates a case where the proportion of white pixels is $\gamma = 0$.}
    \label{fig:image_samples}
\end{figure}
%-------------------------------------------------------
%-------------------------------------------------------

The dataset comprises annotated images, each of which is a 5000-by-5000 pixel RGB satellite image with a spatial resolution of 30cm. The images are accompanied by geospatial coordinates and manually annotated polygons for each solar array. These annotations facilitate tasks such as image classification, semantic segmentation and object detection.

Before using this dataset, two main considerations must be taken into account. Firstly, the dataset is not readily prepared for classification tasks. Secondly, it is too large and complex to be fed raw into any quantum algorithm in the NISQ era. Therefore, it is necessary to take a few steps to simplify the problem, reduce the size of the data and make it available for classification tasks. The data preparation pipeline involves the following steps:

\begin{enumerate}
    \item \textit{Convert polygon annotations to segmentation masks.} The dataset includes georeferenced annotations in JSON format, which specify the coordinates of the polygons that comprise the panels. To label each image and extract features from them, it was necessary to convert these polygons to binary segmentation masks where each pixel indicates the presence (white) or absence (black) of solar arrays. The process was carried out using the open-source \texttt{scikit-image} Python library, which offers features for drawing the inner area of a polygon.
    
    \item \textit{Image partitioning.} Working with high-resolution images is demanding in terms of memory usage and computational time. This challenge, already significant for classical hardware, becomes even more pronounced on quantum devices due to limited quantum resources, such as the number of available qubits and allowable circuit depth. To address these constraints, each 5000-by-5000-pixel image is divided into 400 smaller images, each measuring 250-by-250 pixels. While all original high-resolution images contain panels, cropping small sections of these images produces samples without panels, enabling the creation of a dataset with both positive samples (images containing panels) and negative samples (images without panels).
    
    \item \textit{Sampling and rebalance classes.} After slicing the original 463 images, we obtained 185,200 smaller images, of which only 10,863 were positive instances, representing 5.768\% of the total. Defining a ``positive instance'' for each image is not straightforward, as this information is not directly available in the dataset. However, using the image mask, we can establish a local criterion to assign class labels. In this work, an image’s class label is defined as follows

    \begin{equation}
        \hat{y}\in\left\{\begin{matrix}
        +1, & \mathrm{if} \: \gamma>\varepsilon  \\ 
        -1, & \mathrm{if} \:\gamma=0,
        \end{matrix}\right.
        \label{eq:1}
    \end{equation}
where $\gamma=\frac{1}{N} \sum_{i,j} M_{ij}$ represents the proportion of positive (white) pixels in each image mask $M_{ij} \in \{0,1\}$, $\varepsilon$ is a threshold set to exclude ``edge cases'' and $N$ is the total number of pixels in each image. Note that if an image’s mask contains a proportion of white pixels between zero and the threshold $\varepsilon$, it is excluded from the dataset.
    
    Additionally the dataset's class imbalance was removed ad-hoc by undersampling the majority class. 
    
    \item \textit{Prune the positive cases.} As defined in Eq.~\ref{eq:1}, it is necessary to establish a threshold to filter out non-representative cases. For instance, an image containing only a few pixels of solar arrays may be visually indistinguishable from a negative image, even to the human eye. Expecting a classification model to accurately identify such "edge cases," where the object is barely recognizable, would be unrealistic. Therefore, to avoid training the model on structures that might be confused with background noise, these ambiguous cases are excluded from the dataset.

    The threshold value $\varepsilon$ for filtering was determined based on the distribution of solar array pixels across all cropped images. Specifically, the 15th percentile of the pixel count distribution was selected as the minimum value to retain positive samples. This threshold corresponds to a maximum panel coverage of 0.2\% of the entire cropped image area and balances the goal of retaining sufficient samples while ensuring a high-quality dataset.

\end{enumerate}

A final outcome example of a positive and a negative sample image is shown in Figure~\ref{fig:image_samples}. It is important to note that the dataset is not yet fully prepared for use and encoding in a quantum circuit. A step of thorough feature extraction and dimensionality reduction is needed, as will be seen.

%----------------------FIGURE 2-------------------------
%-------------------------------------------------------
%-------------------------------------------------------
%-------------------------------------------------------

%----------------------FIGURE 3-------------------------
%-------------------------------------------------------
\begin{figure*}[t!]
    \centering
    \includegraphics[width=1.0\textwidth]{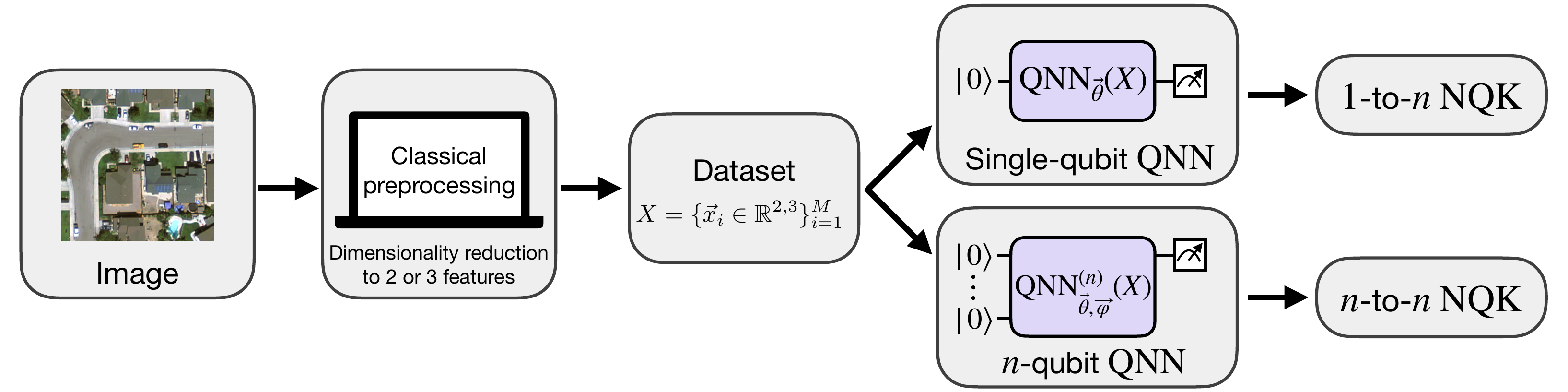}
    \caption{Diagram illustrating the complete architecture, from the input image to the construction of the NQK. Starting with large, complex images, a dataset $X$ of $M$ data points is generated, consisting of 2- or 3-dimensional points, through classical preprocessing. The upper branch represents the $1$-to-$n$ NQK construction, while the lower branch illustrates the $n$-to-$n$ NQK, which is built using an $n$-qubit QNN. This QNN is trained iteratively to ensure scalability. $\vec{\theta}$ and $\vec{\varphi}$ denote the trainable parameters.}
    \label{fig:Pipeline}
\end{figure*}
%-------------------------------------------------------
%-------------------------------------------------------
\newpage
\subsection{Feature engineering}\label{sub:feat_eng}
Since the input to the problem consists of geospatial images, and the purpose of this work is to study QML algorithms designed for tabular data, it becomes necessary to extract tabular features from the images. In recent years, widespread methods for extracting features from high-dimensional (i.e. high-resolution) images have relied on the use of pre-trained classification CNNs (e.g., ResNet, EfficientNet, etc.) as well as the use of CAEs~\cite{cae} . Preliminary tests indicated that solving the PV identification problem classically using pre-trained CNNs was not viable, as benchmarking the extracted embeddings revealed low performance. This observation led to the conclusion that general-purpose multiclass CNNs trained on ImageNet may be unsuitable for detecting small objects in high-resolution aerial or satellite images, which typically feature diverse geometric structures and geospatial patterns. On the other hand, CAE architectures capable of learning to reconstruct the input image are often considered for feature extraction and dimensionality reduction of images before performing other downstream tasks. In both cases, the so-called \textit{bottleneck features} (also known as the \textit{latent vectors}) represent a global summary of the input image.

Given the specificity of the small structure identification problem in this work, we decide to approach the classification problem in a different way: from a ``general to specific approach'' to a ``specific to general approach''. While identifying solar panels in an image is generally considered a global classification task, a more specific task is to identify the exact silhouette of the object in the image. Semantic segmentation techniques are commonly used to achieve this, using CAE architectures such as the well-known U-Net~\cite{ronneberger2015unet}. Thus, it was determined that training a U-Net architecture from scratch using the images and their segmentation masks would yield optimal features for small object identification. This approach was previously proposed in Ref.~\cite{semantic-seg-features} for classifying small lesions in medical images using features extracted from the bottleneck of semantic segmentation models. 

The segmentation features used for the binary classification task were extracted from a slightly modified U-Net architecture, which is further described in Appendix~\ref{ap:unet}. This method was found to be particularly efficient both as a feature extraction method and dimensionality reduction strategy in order to smartly pave the way to the usage of quantum algorithms that can be tackled on NISQ computers. The network was trained from scratch with a set of training images. To extract the latent vector features, only the encoder part up to the bottleneck layer is used for each image. 

The encoder produces 64-sized vectors as output from the bottleneck dense layer. Depending on the size of the parameterized circuit, these features can be further reduced through dimensionality reduction techniques. Once the latent features have been obtained, $z$-score normalization is applied to them prior to the application of dimensionality reduction through PCA. The number of components $p$ for PCA is selected as a hyperparameter of the model that depends on the specific requirements of the circuit's encoding. Before embedding the features into the quantum circuit, their values are scaled to fit within the range of \([-1, 1]\). Finally, the circuit yields the prediction \(\hat{y}\), that is either +1 (positive class) or -1 (negative class).

It is important to mention that a large variety of alternative feature extractions techniques might also have been considered, such as: a) classical computer vision features extraction techniques applied at image level (e.g. SIFT, HOG, etc.), b) object-based image analysis techniques based on the sequential application of unsupervised segmentation methods (such as Quickshift, Felzenszwalb, SNIC, SLIC, etc.) followed by extraction of features related to shapes, colors, and other useful image patterns (e.g. GLCM, Hough transform, etc) , c) the application of other deep learning techniques adopting a neural network for extracting features in a more end-to-end, task oriented fashion (e.g.,  VGGNet, ResNet, EfficientNet, Vision Transformers, etc.). While some of them may also result to be effective for the task at hand, a comparison of alternative feature extraction methods is not implemented in this work that instead focuses on the application of the proposed hybrid QML algorithms, and their variants, to the selected complex real dataset use case.

\section{Methodology}
\label{sec:methodology}
Classical pre-processing serves as a vital step in handling highly intricate real-world datasets. Within our pre-processing framework detailed in Section~\ref{sec:dataset}, PCA stands out as a key technique which inherently introduces nonlinearities to the model. This way, the choice of the number of principal components $p$ becomes a crucial hyperparameter of the model.

In our study, we focus on $p=2$ and $p=3$ features since we are using qubits for the quantum part. Additionally, despite retaining more data information, considering a larger number of principal components does not guarantee improved performance, as shown in the Appendix.

After reducing the data dimensionality, we combine two well-known QML architectures, namely QNNs and EQKs, to create a more robust model termed NQKs.

%----------------------FIGURE 3-------------------------
%-------------------------------------------------------
\begin{figure*}[t!]
    \centering
    \includegraphics[width=\textwidth]{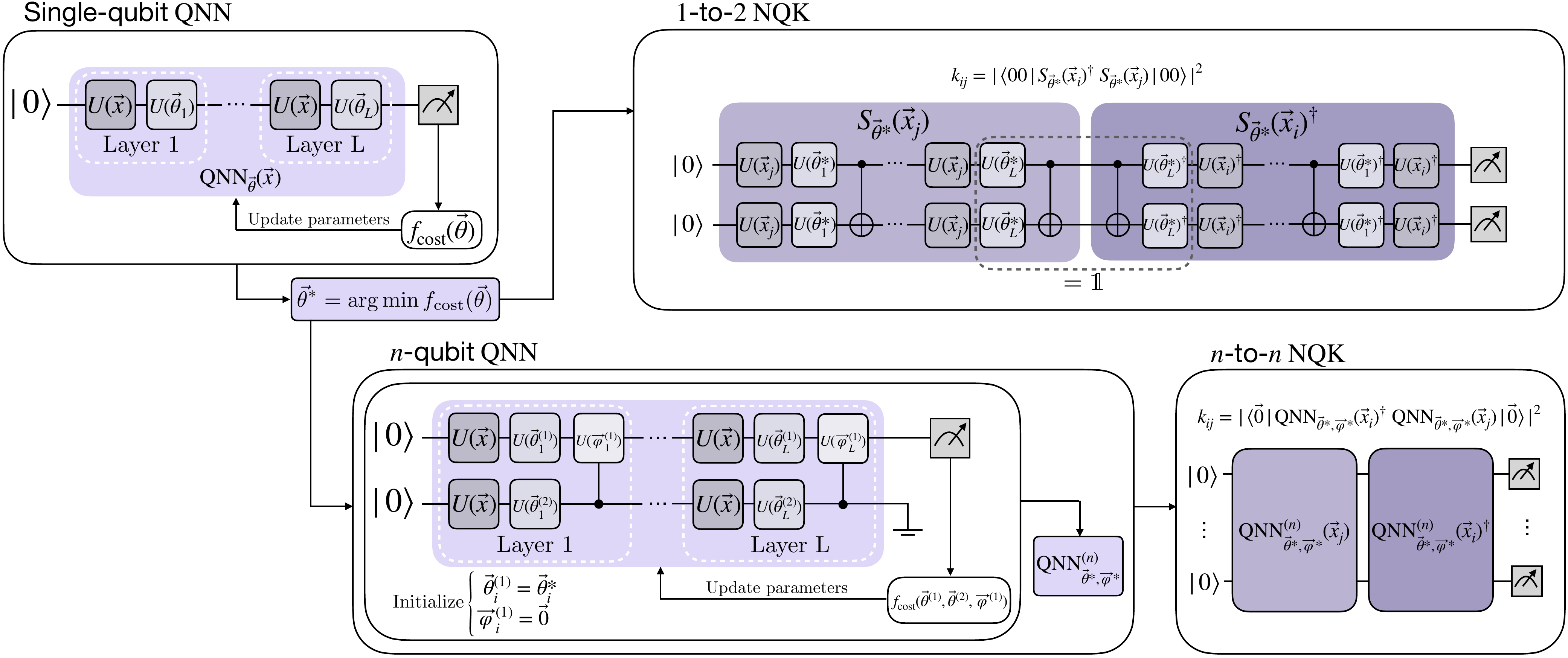}
    \caption{Detailed NQK construction process. The diagram begins with the training of a single-qubit QNN, from which the optimal parameters $\vec{\theta}^*$ are obtained. Using these parameters, two types of NQKs can be constructed. The first is the $1$-to-$n$ NQK, which is shown in the upper part of the figure, illustrating the specific case where $n=2$. The second is the $n$-to-$n$ NQK, where we iteratively add qubits to construct an $n$-qubit QNN. The first step in constructing the 2-qubit QNN is shown in the lower-left part of the figure. This $n$-qubit QNN is then used directly as the embedding to create the $n$-to-$n$ NQK, as shown in the lower-right part.} 
    \label{fig:quantum_part}
\end{figure*}
%-------------------------------------------------------
%-------------------------------------------------------
\subsection{Data re-uploading QNN}
We adopt the data re-uploading architecture described in Refs.~\cite{P_rez_Salinas_2020, P_rez_Salinas_2021} for our QNN. For a single-qubit, this is defined by
\begin{equation}\label{reuploading_QNN}
    \mathrm{QNN}_{\vec{\theta}}(\vec{x})\equiv \prod_{l=1}^L U(\vec{\theta}_l)\;U(\vec{x})= U(\vec{\theta}_L )\;U(\vec{x}) \dots U(\vec{\theta}_1)\;U(\vec{x}).
\end{equation}
This structure alternates parametrized gates $U(\vec{\theta}_l)$ with encoding gates $U(\vec{x})$.  Here, $U$ represents a generic $\mathrm{SU}(2)$ unitary, parameterized by three angles. The number of free parameters is $3\times L$, where $L$ denotes the number of layers, contained within the vector $\vec{\theta}=\{\vec{\theta}_1,...,\vec{\theta}_L\}$. With a single unitary we can encode up to $p=3$ features. For data points with 
$p=2$ features, we set the third angle of the encoding gate to 0. This architecture yields highly expressive models, as demonstrated by the analyzed Fourier frequencies of the functions generated by the model \cite{data_encoding,multi_fourier}. Remarkably, even with a single qubit, one can perform complex classification tasks \cite{Tapia_2023,single_qubit}. 

The optimal model parameters $\vec{\theta}^*=\{\vec{\theta}^*_1,...,\vec{\theta}^*_L\}$ are defined as
\begin{equation}
    \vec{\theta}^*=\arg\min f_{\mathrm{cost}}(\vec{\theta}).
\end{equation}
The cost function we use is the fidelity cost function
\begin{equation}\label{cost_function}
    f_\mathrm{cost}(\vec{\theta})=\frac{1}{M}\sum_{i=1}^M \Bigl(1-|\langle \phi^i_l|\phi(\vec{\theta},\vec{x}_i)\rangle|^2\Bigr),
\end{equation}
where $|\phi_l^i\rangle$ denotes the correct label state for data point $\vec{x}_i$ (which are chosen to be either $|0\rangle$ or $|1\rangle$) and $|\phi(\vec{\theta},\vec{x}_i)\rangle=\mathrm{QNN}_{\vec{\theta}}(\vec{x}_i)|0\rangle$. Once it is trained, the single-qubit QNN assigns labels according to the decision rule
\begin{equation}
    \hat{y}[\vec{x}_t]=\mathrm{sign}\Bigl(\mathrm{tr}(|0\rangle\langle 0|\;\rho_{\vec{\theta}^*}(\vec{x}_t))-1/2\Bigr)\in\{-1, +1\},
\end{equation}
where $\rho_{\vec{\theta}^*}(\vec{x}_t)\equiv \mathrm{QNN}_{\vec{\theta}^*}(\vec{x}_t) |0\rangle\langle 0| \mathrm{QNN}_{\vec{\theta}^*}(\vec{x}_t)^\dagger$. 

Starting from the single-qubit QNN, we can scale this up by adopting the iterative training strategy outlined in Ref.~\cite{rodriguezgrasa2024training}, where the construction of an $n$-qubit QNN proceeds iteratively. This method gradually adds qubits, ensuring that the performance of the $n$-qubit QNN equals or surpasses that of the $(n-1)$-qubit counterpart.

This iterative construction is based on considering a local measurement on the first qubit. Thus, when adding an additional \(n\)th qubit, if we do so in a way that it is initially decoupled from the previous \(n-1\) qubits and initialize the training with the parameters from the previous step, we can ensure that adding this new qubit can only further reduce the value of the cost function. To achieve this decoupling between qubits, all that needs to be done is to initialize the parameters of the controlled rotations that couple the new qubit to the previous ones to 0, resulting in an identity operation.

To implement this strategy, we start by training a single-qubit QNN to determine the optimal parameters $\vec{\theta}^{(1)*}$. Subsequently, we expand the network to include a second qubit, constructing the following architecture
\begin{equation}
    \mathrm{QNN}^{(2)}_{\vec{\theta},\vec{\varphi}}(\vec{x})=\prod_{l=1}^L \Bigg(\mathrm{CU}^1_{2}(\vec{\varphi}_l^{(1)})\Big( U(\vec{\theta}^{(1)}_l)\otimes U(\vec{\theta}^{(2)}_l)\Big)\;U(\vec{x})^{\otimes n}\Bigg).
\end{equation}
Here, $\vec{\theta}^{(1)}_l$ is initialized with $\vec{\theta}^{(1)*}_l$ and $\vec{\varphi}^{(1)}_l$ is set to zero for each $l\in [1,L]$ as depicted in Figure \ref{fig:quantum_part}. Optimization encompasses all parameters, and the optimal values are then used to initialize the $3$-qubit QNN, following the same procedure. Thus, we can scale this up to construct $\mathrm{QNN}^{(n)}_{\vec{\theta},\vec{\varphi}}(\vec{x})$, where $\vec{\theta}$ contains all the single-qubit gate trainable parameters and $\vec{\varphi}$ the trainable parameters of the entangling gates.

In this process, the cost function is the same as in Eq.~\ref{cost_function}, but including now dependency on $\vec{\varphi}$, and the decision rule is expressed as 
\begin{equation}
    \hat{y}[\vec{x}_t]=\mathrm{sign}\left(\mathrm{tr}\left(|0\rangle\langle 0|\otimes\mathds{1}^{(n-1)}\;\rho^{(n)}_{\vec{\theta}^*,\vec{\varphi}^*}(\vec{x}_t)\right)-1/2\right),
\end{equation}
where $\rho^{(n)}_{\vec{\theta}^*,\vec{\varphi}^*}(\vec{x}_t)\equiv \mathrm{QNN}^{(n)}_{\vec{\theta}^*,\vec{\varphi}^*}(\vec{x}_t) |\vec{0}\rangle\langle\vec{0}| \mathrm{QNN}^{(n)}_{\vec{\theta}^*,\vec{\varphi}^*}(\vec{x}_t)^\dagger$, is the quantum state which comes from applying the trained QNN to the $|\vec{0}\rangle$ state.

\subsection{Neural quantum kernels }
As introduced in Ref.~\cite{rodriguezgrasa2024training}, Neural quantum kernels (NQKs) are parameterized EQKs whose parameters are obtained from the training of a QNN. Generally, we denote $n-$to$-n\cdot m$ NQK to signify that a $n-$qubit QNN was used to construct a $n\cdot m-$qubit EQK.

Considering some classical data point $\vec{x}$ and a quantum embedding $S(\vec{x})$, EQKs are constructed as follows
\begin{equation}\label{kernel}
    k(\vec{x}_i,\vec{x}_j)
    %=|\langle \phi(\boldsymbol{x}_i)|\phi(\boldsymbol{x}_j)\rangle |^2
    =|\langle \vec{0}| S(\vec{x}_i)^\dagger\; S(\vec{x}_j)|\vec{0}\rangle|^2,
\end{equation}
where $|\vec{0}\rangle=|0\rangle^{\otimes n}$. However, the optimal embedding varies depending on the problem. Therefore, we allow the embedding to be parameterized, defining a parameterized EQK. For NQKs, the embedding parameters are obtained from the training of a QNN.

As depicted in Figure \ref{fig:quantum_part}, we distinguish between two different NQK architectures: the $1$-to-$n$ and the $n$-to-$n$. In the $1$-to-$n$ approach, a single-qubit QNN is trained, and the obtained parameters are used to construct an $n$-qubit EQK. In the $n$-to-$n$ approach, the single-qubit QNN is scaled to an $n$-qubit QNN, and this architecture is directly used to construct an $n$-qubit EQK. By combining these approaches, one could build the general $n$-to-$n \cdot m$ NQK model.

To construct a $1$-to-$n$ NQK, we proceed as follows: we replicate the data re-uploading QNN structure across the $n$ qubits, using the parameters $\vec{\theta}^*$ obtained from the QNN training as the arguments for the parameterized unitaries. Between layers, we implement a cascade of CNOT gates between nearest neighbor qubits, though other configurations could also be chosen. The precise expression of such embedding is
\begin{equation}
    S_{\vec{\theta}^*}(\vec{x})=\prod_{l=1}^{L}\left(\left(\prod_{s=1}^{n-1}\mathrm{CNOT}^{s+1}_s\right)\;U(\vec{\theta}^*_l)^{\otimes n}\;U(\vec{x}_j)^{\otimes n}\right),
\end{equation}
where we use the convention $\prod_{i=1}^n A_i=A_n\dots A_1$, and in $\mathrm{CNOT}^{s+1}_s$, the subscript and the superscript denote the control and the target qubits, respectively. The detailed construction of the embedding for the $1$-to-$2$ case is shown in Figure \ref{fig:quantum_part}.

On the other hand, the $n$-to-$n$ case requires only taking the trained $n$-qubit QNN directly as the embedding
\begin{equation}
    S_{\vec{\theta}^*,\vec{\varphi}^*}(\vec{x})=\mathrm{QNN}_{\vec{\theta}^*,\vec{\varphi}^*}^{(n)}(\vec{x}).
\end{equation}

Once the quantum embeddings have been define, we construct the kernel matrix $K$ with entries $k_{ij}=k(\vec{x}_i,\vec{x}_j)$. This matrix serves as input for a SVM, which solves a convex optimization problem to find optimal $\vec{\alpha}$ parameters defining the decision rule
\begin{equation}
    \hat{y}[\vec{x}_t] = \mathrm{sign}\left(\sum_{i=1}^M \alpha_i\;\hat{y}_i\; \frac{k(\vec{x}_t, \vec{x}_i) + 1}{2}\right),
\end{equation}
where $\{\vec{x}_i\}_{i=1}^M$ denotes training points from set $X$.

Figure \ref{fig:quantum_part} illustrates the entire process of constructing NQKs for both the $1$-to-$n$ and $n$-to-$n$ cases. The process begins with the training of a single-qubit neural network (top left), where we obtain the optimal parameters \(\vec{\theta}^*\). Once we have these parameters, two construction approaches are possible. The $1$-to-$n$ approach, exemplified for the case $n=2$ in the top right, involves replicating the architecture of the trained single-qubit QNN and adding CNOT gates between the two qubits in each layer. This results in a 2-qubit EQK, referred to as a 1-to-2 NQK. The $n$-to-$n$ approach, described at the bottom of the figure, requires constructing an $n$-qubit QNN through a scalable iterative process explained earlier in this section. Once the QNN is built, it is used as an embedding to generate an $n$-qubit EQK, as shown in the bottom right.

In the construction of NQKs, training the QNN integrates information about the classification problem into the EQK, resulting in a more robust QML model. Indeed, in the following section, we demonstrate that even a QNN that is not perfectly trained can still effectively determine parameters to construct an NQK that performs remarkably well.

\begin{figure*}[t!]
    \centering
    \includegraphics[width=1.0\textwidth]{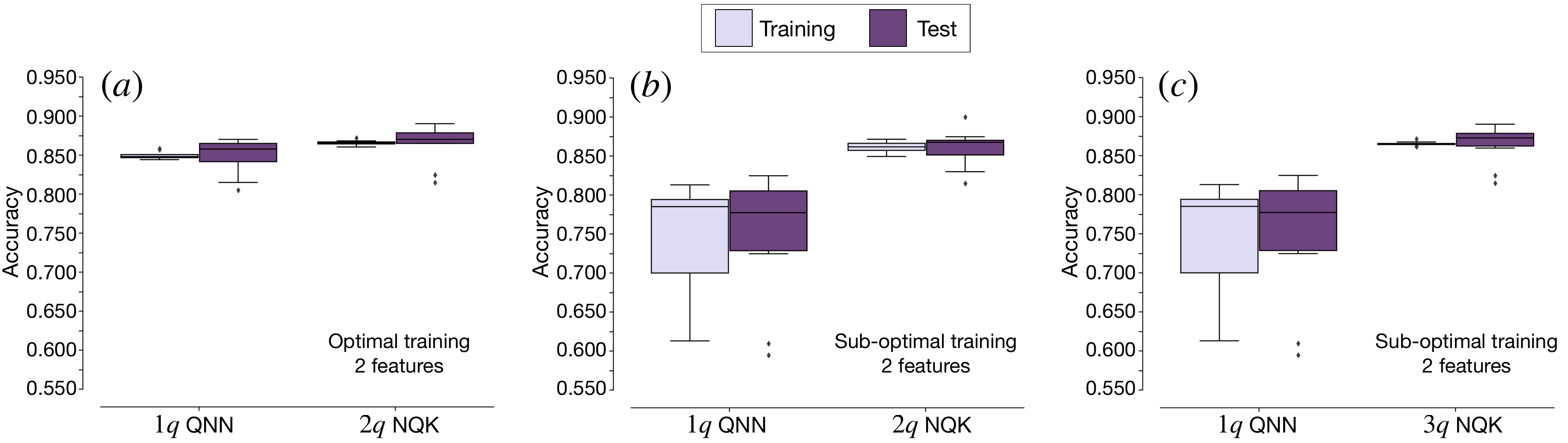}
    \caption{Numerical results for the $1$-to-$n$ NQKs approach. $1q$, $2q$, and $3q$ refer to one-qubit, two-qubit, and three-qubit constructions, respectively. In $(a)$ and $(b)$, we consider the $1$-to-$2$ NQK, while in $(c)$ we examine the $1$-to-$3$ NQK. The results are displayed for $p=2$ features. The training and test accuracies are drawn from the 10-fold process using the described ``$1$-to-$n$ set'' of 2000 samples. In the box plots, the upper and lower parts of the boxes represent the 25th and 75th percentiles of the accuracies obtained by evaluating the accuracy of the classifiers using 10-fold. The horizontal line inside the box indicates the median, while the whiskers extend from the box to the most distant evaluation score that lies within 1.5 times the interquartile range (IQR) from the box. Score values beyond the whiskers are considered outliers.}
    \label{fig:Numerical}
\end{figure*}
\section{Numerical results and analysis}
\label{sec:results}
Once introducing both, the real-world classification problem and the methodologies, we demonstrate that we can achieve performance levels ranging between 85-90\%, which is remarkable given the complexity of the classification task. In the following section, we present numerical results showcasing the performance of the described models. Additionally, we provide a classical benchmark for the classification of this dataset, aiming to be as rigorous as possible in evaluating and comparing the performance of the different models. 

For each experiment, a different split of the dataset is used to ensure there is no data leakage: 
\begin{itemize}
    \item \textbf{U-Net training set}: this set of 10,895 images is used just to train the U-Net architecture from scratch to generate the embeddings.

    \item \textbf{U-Net test set}: this set comprising 6,630 images is used for inference on the U-Net to extract image latent features. This approach ensures that there is no data leakage on the extracted embeddings, since the mask is not used.

    \item \textbf{$1$-to-$n$ set}: it is a 2000-sample random subset from the U-Net test set to be used in both the $1$-to-$n$ NQK experiments and the classical benchmarking. It is used by performing the stratified 10-fold strategy to get robust test results.

    \item \textbf{$n$-to-$n$ set}: random subsets from the U-Net test set are selected to perform $n$-to-$n$ NQK experiments. We consider a total of 700 samples, which are divided into 500 training points and 200 test points.
    
\end{itemize}

Please refer to Appendix~\ref{ap:dataset_splits} for further details on the approximation carried out in this work to avoid leakage through dataset splits.

\subsection{$1$-to-$n$ NQK}
For these experiments, we used the ``$1$-to-$n$ set'' comprising 2000 images. The $k$-fold method was performed with $k=10$ to obtain robust and reliable evaluation metrics. On each of the 10 iterations, 9 folds of the dataset were used for training (1800 images), and 1 fold was used for testing (200 images). This approach yields the results illustrated in Figure~\ref{fig:Numerical}, which depicts the training and test accuracies for the single-qubit QNN and the NQK. This visualization highlights that models with fewer or no outliers and narrower boxes are more robust, as their performance remains more consistent across different evaluation folds.

We examine both $p=2$ and $p=3$ features but primarily focused on $p=2$, as using three features  performs less effectively with this model. The results for $p=3$ are presented in Appendix~\ref{ap:numerical_experiments}. For the number of qubits in the EQK, we considered $n=2$ and $n=3$, corresponding to $1$-to-$2$ and $1$-to-$3$ NQKs, respectively. For this classification task, increasing the number of qubits does not enhance performance. Notably, we achieved high classification accuracies even with a small number of qubits.

We distinguish between optimal and sub-optimal QNN training scenarios, both employing an Adam optimizer. For optimal QNN training, the QNN is allowed to converge over 10 epochs with a learning rate of 0.01. In contrast, for sub-optimal QNN training, the process is truncated after 2 epochs to prevent convergence of the loss function, and a lower learning rate of 0.001 was used. This differentiation aims to demonstrate the robustness of NQKs. We observe that while better trained QNNs show minimal improvement from introducing NQK, the NQK's performance remains high even when the QNN training is suboptimal. This indicates that only a few training iterations are sufficient to select a suitable embedding for constructing a robust NQK. While optimal QNN training is straightforward for systems with few qubits, the robustness of NQK becomes crucial for larger QML models where optimal training is not guaranteed. In such cases, NQKs could be a candidate to circumvent QNN trainability problems.

In analyzing Figure \ref{fig:Numerical}, we observe that the best results across the three plots reach close to 90\% accuracy, with a mean around 86\%. In Figure \ref{fig:Numerical} $(a)$, where the single-qubit QNN is optimally trained, constructing the 2-qubit NQK offers minimal improvement, although the results with NQK are more concentrated. In Figure \ref{fig:Numerical} $(b)$, with sub-optimal QNN training, QNN accuracies drop to around 75-80\%, while the 2-qubit NQK performs nearly as well as in the optimal training scenario. Additionally, the wider boxes of the QNN indicate a strong dependency on parameter initialization and the image set used, whereas the narrow NQK boxes demonstrate its robustness. Figure \ref{fig:Numerical} $(c)$, corresponding to the sub-optimal scenario with a 3-qubit EQK, shows results similar to the 2-qubit case but more concentrated around the mean. This may be due to the larger Hilbert space enhancing the linear separation. 
%All three plots are generated using the same folds, which is why the sub-optimal 1-qubit QNN appears identical in the middle and right subplots.

\begin{figure*}[t!]
    \centering
    \includegraphics[width=1.0\textwidth]{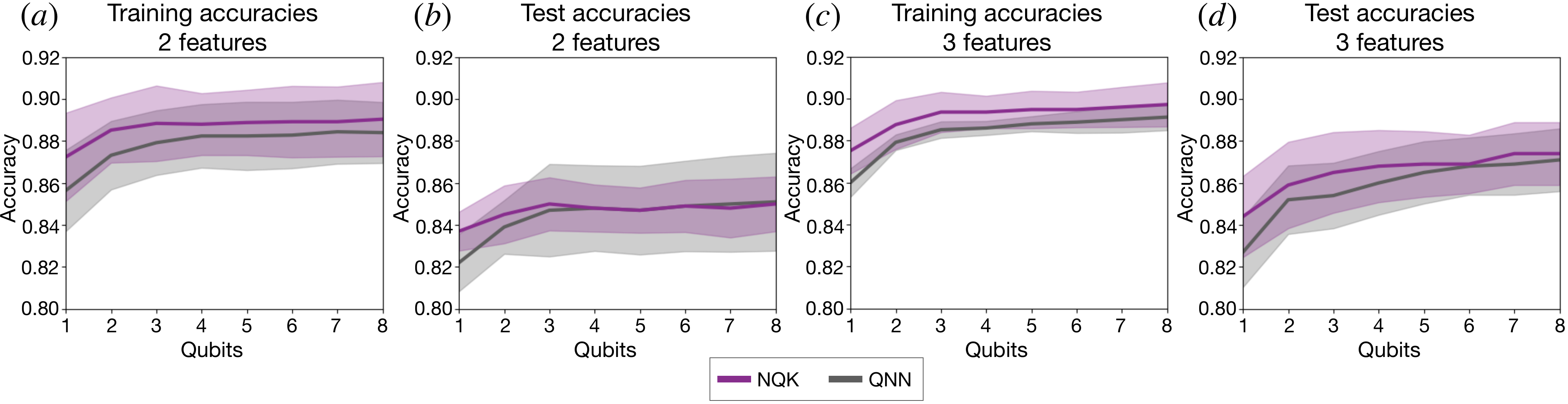}
    \caption{Numerical results for the $n$-to-$n$ architecture, tested using datasets with 500 training points and 200 test points. The experiments show the mean results from five independent trials, with 2 features displayed in panels (a) and (b), and 3 features in panels (c) and (d). The ``scale QNN set'', which is described in the main text, was utilized for these experiments. The figure displays the average training and test accuracies for both the QNN and the corresponding $n$-to-$n$ NQK.}
    \label{fig:scale_QNN}
\end{figure*}
\subsection{$n$-to-$n$ NQK}
In our second approach, instead of constructing an EQK after training a single-qubit QNN, we scale the model up to an $n$-qubit QNN and then derive an $n$-qubit NQK. This scaling is performed iteratively, ensuring that the cost function does not increase when adding qubits, allowing for progressive model improvement.

The results are presented in Figure \ref{fig:scale_QNN}, which shows the mean training and test accuracies along with the standard deviation across 5 independent runs, each using distinct training and test sets. For this analysis, we employed a 6-layer QNN, the Adam optimizer with a learning rate of 0.005, and 10 epochs, scaling the model up to $n=8$ qubits.

The four plots in the figure reveal that accuracies generally increase with the number of qubits, with a more pronounced improvement in training accuracy due to its direct link to the cost function. The standard deviation of test accuracies is naturally larger, reflecting the variability across different subsets.

In Figures \ref{fig:scale_QNN} $(a)$ and $(b)$, results with $p=2$ features are shown, while Figures \ref{fig:scale_QNN} $(c)$ and $(d)$ depict the outcomes with $p=3$ features. Overall, the NQK approach consistently outperforms the QNN architecture across all configurations. Specifically, with $p=3$ features, the average training accuracy nears 90\%, and the mean test accuracy exceeds 86\%. In contrast, with $p=2$ features, the NQK approach achieves average training and test accuracies of over 88\% and 84\%, respectively. However, slight overfitting is observed, as indicated by the higher training accuracies compared to test accuracies, which is expected given the dataset’s complexity.

\subsection{Classical benchmark}

\begin{table}[ht!]
\centering
\begin{tabular}{c|c|c|c}
\hline
\textbf{Model} & \textbf{\# features ($p$)}  & \textbf{Training accuracy} & \textbf{Test accuracy}\\
\hline
\multirow{3}{*}{SVC} & $p=2$ & 86.1$\pm$0.3 & 86.2$\pm$2.5 \\
 & $p=3$ & 88.0$\pm$0.3 & 88.1$\pm$2.7 \\
 & $p=45$ & 89.4$\pm$0.3& 89.4$\pm$2.5\\
\hline
\multirow{3}{*}{Random Forest} 
& $p=2$ & 91.1$\pm$0.3 & 87.2$\pm$2.7 \\
& $p=3$ & 93.0$\pm$0.3 & 87.6$\pm$2.0 \\
& $p=10$ & 96.9$\pm$0.2 & 88.3$\pm$2.8\\
\hline
\multirow{2}{*}{$1$-to-$2$ NQK} 
& $p=2$ & 86.6$\pm$0.3 & 86.5$\pm$2.5  \\
& $p=3$ & 81.7$\pm$1.8 & 81.1$\pm$2.9 \\
\hline
\multirow{2}{*}{$1$-to-$3$ NQK} 
& $p=2$ & 86.6$\pm$0.3 & 86.5$\pm$2.6  \\
& $p=3$ & 85.9$\pm$0.6 & 85.6$\pm$2.7 \\
\hline
\end{tabular}

\caption{Mean training and test accuracies for two classical classification methods are presented. All experiments employ the same 10-fold setup used in the NQK experiments, with 1800 training points and 200 test points. The results from the NQKs come from the optimal training of the QNNs.}
\label{tab:classical_benchmark}
\end{table}

Until now, our investigation has revolved around a hybrid architecture combining classical preprocessing with two distinct QML methodologies. Presently, we delve into classical approaches as substitutes for the quantum components to benchmark our models. In order to accomplish this, we use the preprocessing framework outlined in Section \ref{sec:dataset} and then perform the same 10-fold approximation for training and evaluation with classical machine learning algorithms. 

The selected models for benchmarking are Support Vector Machine classifier (SVC) and Random Forest classifier. As shown in Table~\ref{tab:classical_benchmark}, three runs are performed for each model. In order to ensure a fair comparison with the NQK models, the number of features is fixed to $p=2$ and $p=3$ in the first two rows for each model. A hyperparameter optimization with randomized search is then performed to yield nearly optimal test results. In the last row for each model, also $p$ is treated as an hyperparameter in the randomized search to determine if it would yield better results than using $p \in \{2,3\}$. The results of our analysis indicate that the optimal value of $p$ for the SVC is $p=45$, while for the Random Forest it is $p=10$. The values for the remaining hyperparameters for each model can be found in Appendix~\ref{ap:hyperparams}.

\section{Discussion and limitations}\label{discussion}
This work highlights some conclusions from previous studies while also paving the way for new insights. As observed, the feature selection and classical preprocessing pipeline significantly influences the overall performance of the model, a finding consistent with prior literature \cite{image_quantum, pqgates}. Our results further emphasize that, with effective classical preprocessing, quantum models can be competitive in addressing real-world classification problems. Notably, these results are achievable with a limited number of qubits, making such models potentially quantum-inspired, as they can be directly simulated on classical computers, which opens the door to the development of new classical models.

However, as shown in our classical benchmark models, the findings also underline that classical models remain highly competitive in terms of performance. This suggests that the benefits of using quantum or quantum-inspired models may not lie solely in accuracy but in other metrics such as a reduced number of parameters or improved generalization capabilities, which merit further investigation.

Despite these promising insights, this work has limitations that should be addressed in future research. First, the model is restricted to a very small number of features. As observed, the straightforward method of adding features beyond  proves ineffective, highlighting the necessity of exploring alternative approaches. A promising direction could involve leveraging qudits, which offer more degrees of freedom and could enable the encoding of additional features through single-qudit gates \cite{qudits, Roca_Jerat_2024}.

Another significant limitation is the inherent hardware noise in quantum processing units. While simulations have been used here as an initial proof of concept to showcase the competitiveness of quantum models, future studies should evaluate the effects of noise and the robustness of the models under realistic hardware conditions. Furthermore, if the addition of more features requires additional quantum gates, it becomes even more crucial to study the effects of hardware noise, as more gates naturally introduce more noise. Additionally, if more features necessitate the use of more qubits, the current limitations of quantum hardware size must be carefully considered.

\section{Conclusions and future work}\label{sec:conclusion}
This work demonstrates that quantum models, when combined with effective classical preprocessing, can achieve competitive performance on a real-world binary classification task. Our results show that quantum models based on NQKs achieve accuracies close to 90\%, rivaling well-optimized classical methods. We explored two NQK approaches: the \(1\)-to-\(n\) method, which uses a single-qubit QNN and proves particularly robust, and the \(n\)-to-\(n\) method, which improves accuracy as the number of qubits increases, addressing the scalability challenges commonly discussed in quantum machine learning.

Although quantum models show promise, classical models remain highly competitive in terms of accuracy, suggesting that the real advantages of quantum or quantum-inspired methods may lie in metrics beyond raw performance. Future work should focus on scaling these models to more qubits, addressing noise in quantum hardware, and exploring the potential of qudits for handling larger feature sets. This research opens the door for further exploration of quantum techniques in real-world industrial applications.

\section{Acknowledgements}
The authors would like to thank A. Benítez-Buenache for his helpful comments during the writing of the manuscript. This work was supported by the Spanish Ministry of Science and Innovation under the Recovery, Transformation and Resilience Plan (CUCO, MIG-20211005). PR and MS acknowledge support from HORIZON-CL4- 2022-QUANTUM01-SGA project 101113946 OpenSuperQPlus100 of the EU Flagship on Quantum Technologies, the Spanish Ramón y Cajal Grant RYC-2020-030503-I, project Grant No. PID2021-125823NA-I00 funded by MCIN/AEI/10.13039/501100011033 and by “ERDF A way of making Europe” and “ERDF Invest in your Future”, and from the IKUR Strategy under the collaboration agreement between Ikerbasque Foundation and BCAM on behalf of the Department of Education of the Basque Government. This project has also received support from the Spanish Ministry for Digital Transformation and of Civil Service of the Spanish Government through the QUANTUM ENIA project call - Quantum Spain, EU through the Recovery, Transformation and Resilience Plan – NextGenerationEU within the framework of the Digital Spain 2026. We acknowledge funding from Basque Government through Grant No. IT1470-22 and through the ELKARTEK program, project "KUBIT - Kuantikaren Berrikuntzarako Ikasketa Teknologikoa" (KK-2024/00105).

\clearpage
\bibliography{main}
\clearpage

\onecolumngrid
\appendix
\counterwithin{figure}{section}

\section{Modified U-Net architecture}\label{ap:unet}

The classical architecture chosen for semantic segmentation purposes was the well-known U-Net~\cite{ronneberger2015unet} due to its simplicity and adaptability to the PV plants use case. However, the architecture has been slightly modified to extract one-dimensional bottleneck features to be used in classification tasks, as the main goal was not to solve a semantic segmentation problem.

\begin{figure*}[h!]
    \centering
    \includegraphics[width=1.0\textwidth]{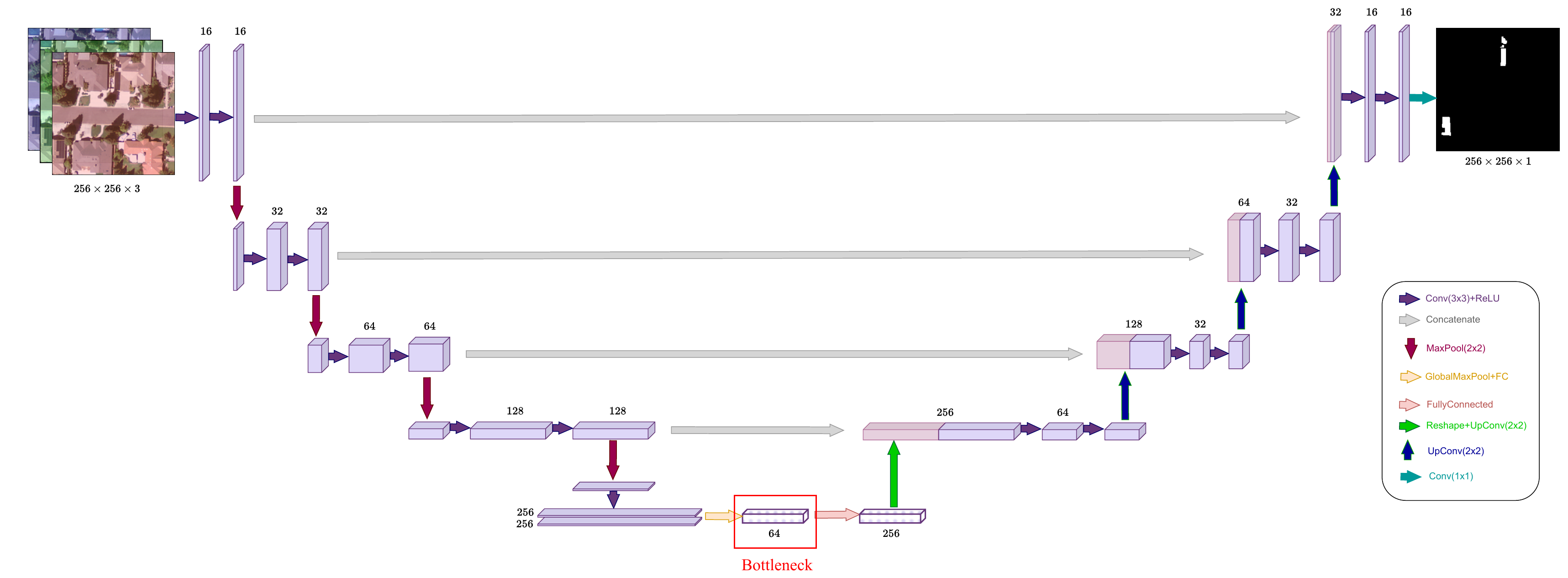}
    \caption{Modified U-Net architecture used to extract bottleneck features}
    \label{fig:bottleneck}
\end{figure*}

As shown in Figure~\ref{fig:bottleneck}, the principal changes in the architecture are:
\begin{itemize}
    \item New added fully-connected layers. At the end of the encoder part of the U-Net, a new $64$-sized fully-connected layer is introduced to be used as the \textit{bottleneck layer} to extract features later. These new layers have been trained from scratch along with the rest of the architecture, thus transfer learning is not applied in this case.
    
    \item Global max pooling and reshaping layers. To adapt the new bottleneck layer, at the end of the last convolutional layer in the encoder, a global max pooling layer is applied to get a one-dimensional output. Following the application of the bottleneck layer, a bidimensional feature map of size $16\times16$ is reconstructed by adding a 256-sized fully-connected layer and then reshaping its output. 
\end{itemize}

\clearpage
\section{Dataset splits}
\label{ap:dataset_splits}
Dividing a dataset in splits is a mandatory practice in machine learning experiments to ensure rigorous results and to avoid data leakage, as it could lead to overly optimistic performance estimates and poor generalization of the models to new data. When dealing with hybrid models with multiple steps in the pipeline, special care must be taken.

\begin{figure*}[h!]
    \centering
    \includegraphics[width=0.5\textwidth]{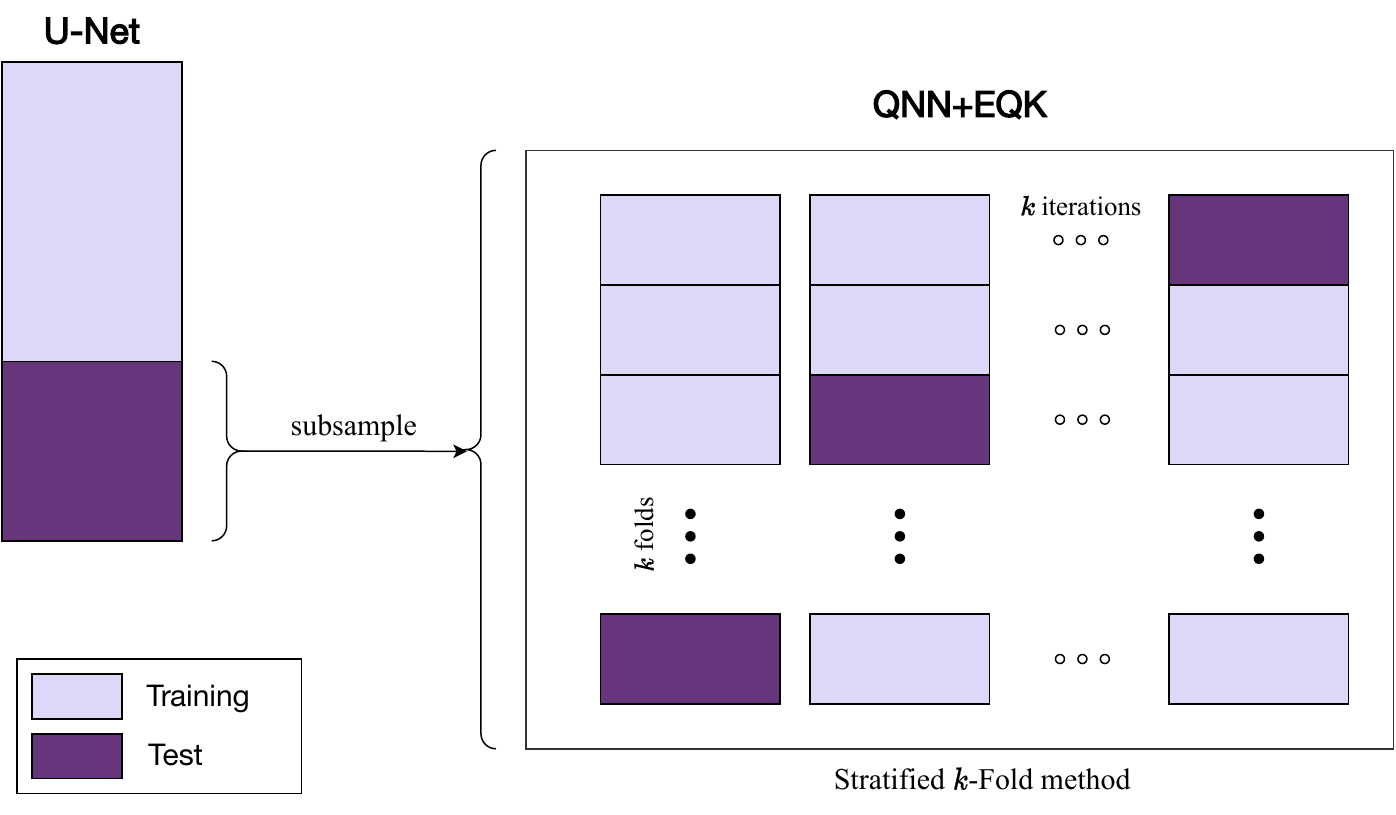}
    \caption{Dataset splits used for each step of the pipeline}
    \label{fig:splits}
\end{figure*}

In semantic segmentation, the label is the binary mask. During the training of the U-Net architecture, some of the mask information from the training samples is being fed to the network through backpropagation. Therefore, the U-Net training samples should not be used in later stages for quantum experiments. Figure~\ref{fig:splits} illustrates the approximation for this work to avoid such phenomenon.

\section{Additional numerical experiments}\label{ap:numerical_experiments}
\subsection{$1$-to-$n$}
\begin{figure*}[h!]
    \centering
    \includegraphics[width=1.0\textwidth]{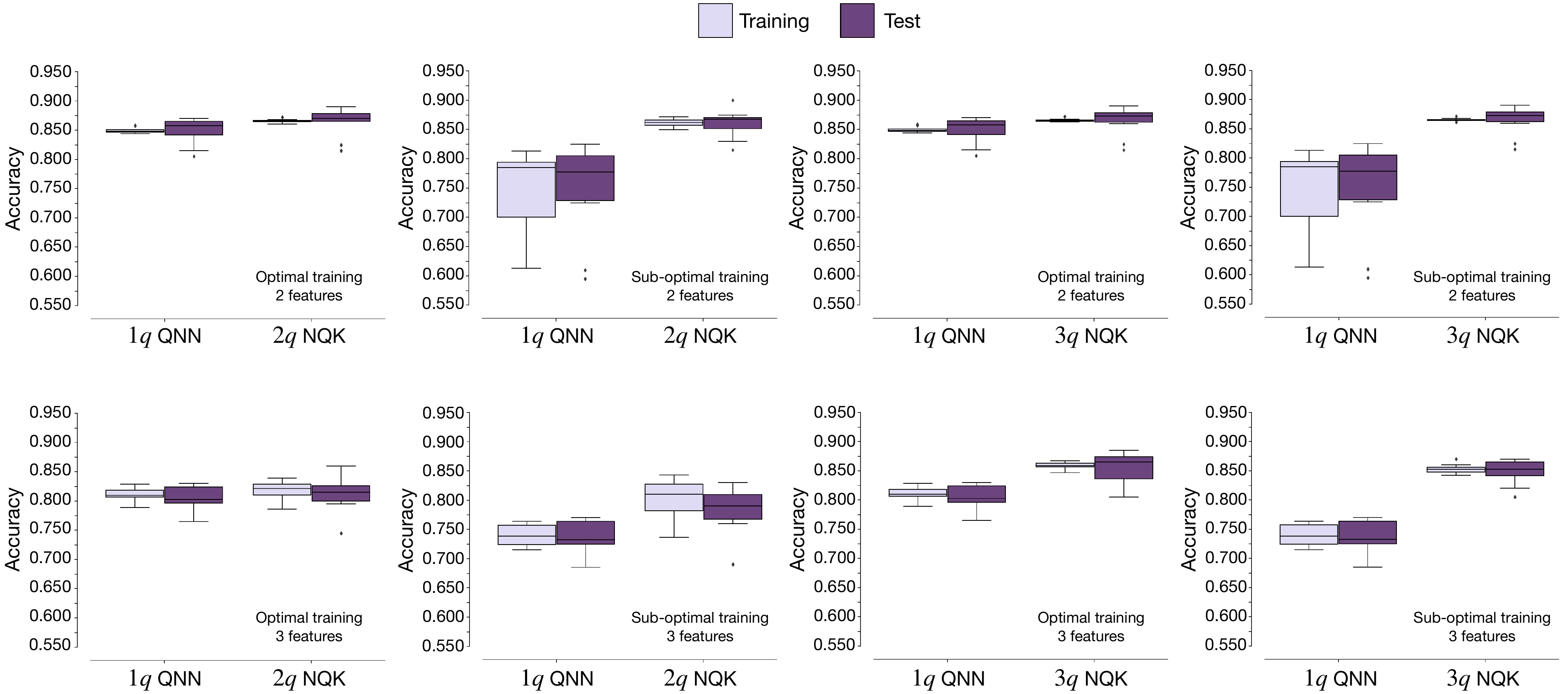}
    \caption{Numerical results for the NQK approach: This figure illustrates the outcomes for all combinations involving $p=2$ and $p=3$ features, optimal and sub-optimal training of the QNN component, and $n=2$ and $n=3$ qubits used in kernel construction.}
    \label{fig:Extra_numerical}
\end{figure*}
Presented here are additional numerical findings not featured in the main text. In Figure \ref{fig:Extra_numerical}, we explore all the $1$-to-$n$ NQK construction scenarios, encompassing $p=2$ and $p=3$ features, optimal and sub-optimal QNN training methods, and $n=2$ and $n=3$ qubits. As discussed in the main text, employing $p=2$ features yields superior performance for this architecture and classification task. We can also observe no improvement when transitioning from $n=2$ to $n=3$ qubits in this context.

\subsection{Different number of features}
As argued in the main text, when we use single-qubit unitaries to encode features, we are limited to encoding only 3 features with a single encoding unitary. For cases where we have $p>3$ features, as proposed in Ref.~\cite{P_rez_Salinas_2020}, we require a number of encoding unitaries equal to $\lceil p/3 \rceil$. For example, if we consider $p=6$ with a data sample $\boldsymbol{x}=(x_1,x_2,x_3,x_4,x_5,x_6)$, instead of using a single encoding unitary, we would need two—namely, $U(x_1,x_2,x_3) \cdot U(x_4,x_5,x_6)$. 

\begin{figure*}[h!]
    \centering
    \includegraphics[width=0.8\textwidth]{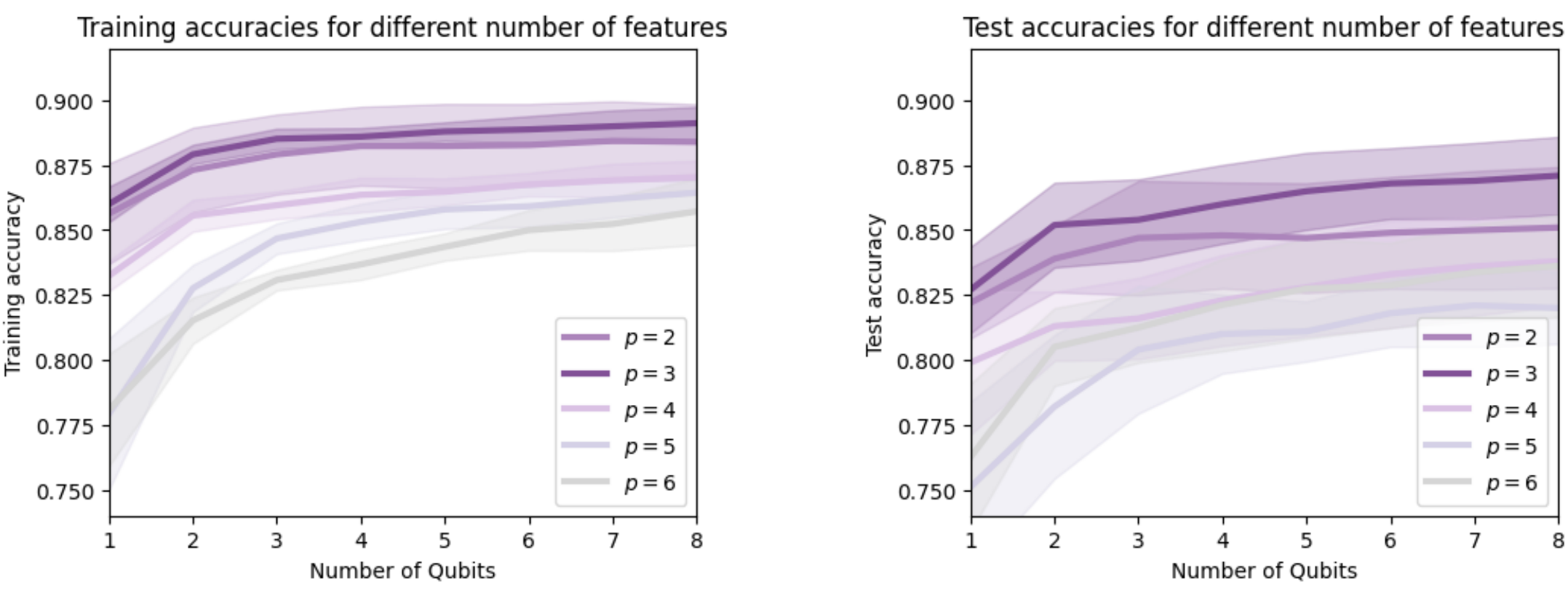}
    \caption{Mean training (left) and test (right) accuracies across 5 dataset samples for different number of features $p$ using the $n$-qubit QNN architecture described in the main text.}
    \label{fig:more_features}
\end{figure*}

However, a single qubit has a limited capacity for processing information, meaning that even with this approach of adding more features, the qubit’s processing ability becomes a bottleneck. Including additional features merely introduces more information than the qubit can handle effectively and increases the circuit depth without necessarily improving model performance. This is evident in Figure~\ref{fig:more_features}, which is intended to clarify why we focus only on $p=2$ and $p=3$. The figure plots the mean training and test accuracies over the 5 dataset samples considered in Figure~\ref{fig:scale_QNN} of the $n$-qubit QNN. 

As we can see, the best performance occurs with $p=3$, followed by $p=2$, where we are still able to use a single encoding unitary. Beyond this, adding more features degrades the results. For training accuracy, performance consistently drops as the number of features increases to $p=4,5,6$, whereas for test accuracy, $p=6$ performs slightly better than $p=5$.

\subsection{Different dimensionality reduction techniques}
In the main text, we used PCA as the feature reduction technique. Here, we present overlaid results for $p=2$ and $p=3$ features obtained with PCA, along with two additional techniques from \texttt{scikit-learn} Python library: Independent Component Analysis (ICA) and Truncated Singular Value Decomposition (SVD). 

\begin{figure*}[h!]
    \centering
    \includegraphics[width=\textwidth]{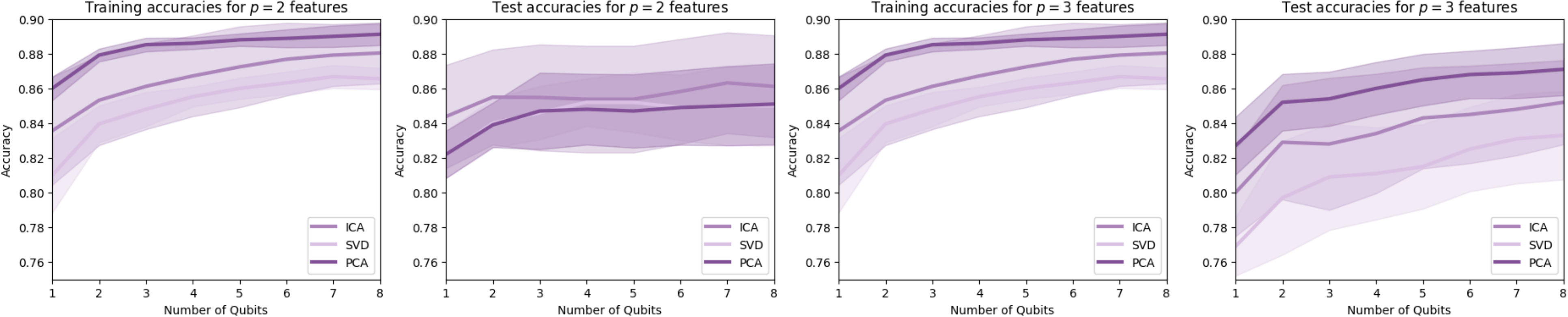}
    \caption{Mean training and test accuracies across five dataset samples from the ``$n$-to-$n$'' set (described in the main text), for $p=2$ and $p=3$ features. The results are presented for three dimensionality reduction techniques: Independent Component Analysis (ICA), Truncated Singular Value Decomposition (SVD), and Principal Component Analysis (PCA), utilizing the $n$-qubit QNN architecture outlined in the main text.}
    \label{fig:dim_reduction}
\end{figure*}

Figure~\ref{fig:dim_reduction} shows the results for the $n$-qubit QNN, presenting the average training and test accuracies across the same five dataset samples used in Figure~\ref{fig:scale_QNN}. As observed, PCA outperforms the other two techniques in most cases, except for the test accuracies with $p=2$ features, where the results across all three techniques are very similar.

\begin{figure*}[h!]
    \centering
    \includegraphics[width=\textwidth]{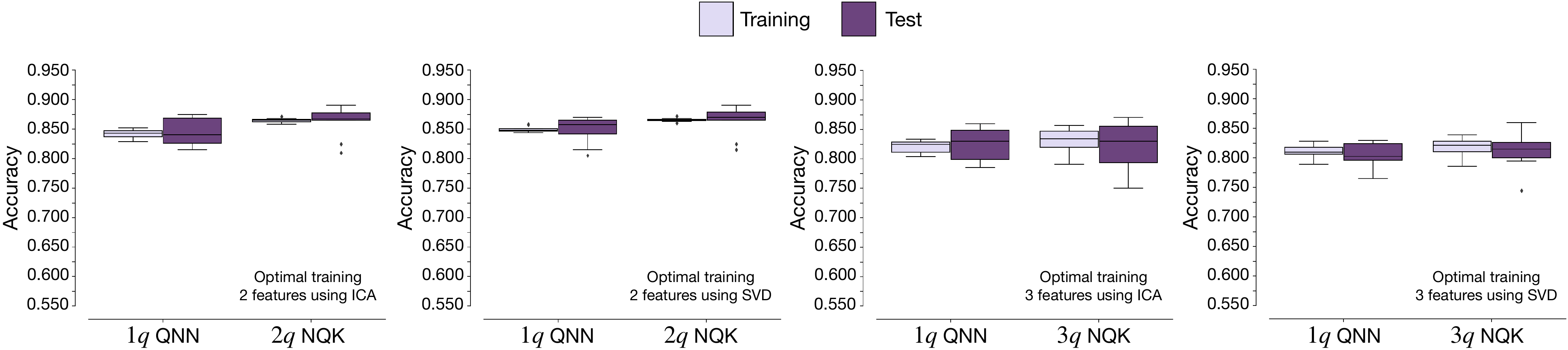}
    \caption{Numerical results for the $1$-to-$2$ NQKs approach with optimal training, incorporating two additional dimensionality reduction techniques: Independent Component Analysis (ICA) and Truncated Singular Value Decomposition (SVD). The results are presented for $p=2$ and $p=3$ features. Training and test accuracies are obtained from a 10-fold cross-validation process using the described ``$1$-to-$n$'' dataset, consisting of 2000 samples. The box plots display the distribution of accuracies, where the upper and lower edges of the boxes represent the 25th and 75th percentiles, respectively. The horizontal line inside the box marks the median accuracy. The whiskers extend from the box to the most extreme evaluation score within 1.5 times the interquartile range (IQR). Any scores beyond the whiskers are considered outliers.}
    \label{fig:NQK_dim_reduction}
\end{figure*}

Figure~\ref{fig:NQK_dim_reduction} shows the results for the $1$-to-$2$ NQK approach with optimal training, using the two additional dimensionality reduction techniques, ICA and SVD. The dataset splits are the same as those used to generate Figure~\ref{fig:Numerical}. As we can see, and when comparing with Figure~\ref{fig:Extra_numerical}, the results using the three dimensionality reduction techniques are quite similar for both $p=2$ and $p=3$ features.

With these additional experiments, we aim to demonstrate that, while the choice of dimensionality reduction method does affect the results, selecting PCA as used in the main experiments is both reasonable and representative for conducting our study.

\section{Hyperparameters for the classical benchmark}\label{ap:hyperparams}

In order to obtain reliable and robust classical results on the dataset to compare with the quantum approaches, a randomized search of the hyperparameters for the selected models (SVC and Random Forest) was performed. It should be noted that the randomised search is not guaranteed to reach the global optimum in the parameter space, but with sufficient iterations it retrieves a satisfactory approximation. The number of iterations set for these experiments was set to 5000.

To ensure a fair comparison, the models were initially tested with the same number of features as the NQK experiments ($p \in \{2,3\}$). Subsequently, the number of components was also set as an hyperparameter on the search, as the original embeddings are of size 64. Tables~\ref{tab:svm_hyperparams} and~\ref{tab:rf_hyperparams} present the values of the hyperparameters identified through the randomized search for each model. \newline

In SVM classifiers, the key hyperparameters are \texttt{kernel}, \texttt{C}, and \texttt{gamma}. The kernel function (e.g., linear, polynomial, RBF) transforms input data to find the optimal hyperplane for classification. The \texttt{C} parameter controls the trade-off between low training error and generalization. The \texttt{gamma} parameter, specific to RBF, polynomial, and sigmoid kernels, defines the influence range of a single training example. In the case of Table~\ref{tab:svm_hyperparams}, even though the parameter \texttt{gamma} is optimized, it has no influence, because the optimal selected kernel is linear in all three cases.

%% SVM HYPERPARAMS %%
\begin{table}[ht!]
\centering
\begin{tabular}{c|c|c|c|c}
\hline
\textbf{Model} & \# features ($p$) & \texttt{kernel} & \texttt{C} & \texttt{gamma} \\
\hline
\multirow{3}{*}{SVC} & $p=2$ & linear & 9.53 & 0.016 \\
 & $p=3$ & linear & 10.09 & 0.079 \\
 & $p=45$ & linear & 10.04 & 0.085 \\
\hline
\end{tabular}

\caption{Optimal values for the hyperparameters of the SVC found by the randomized search}
\label{tab:svm_hyperparams}
\end{table}

In Random Forest classifiers, some of the key hyperparameters are \texttt{n\_estimators}, \texttt{max\_depth}, and \texttt{max\_features}. The \texttt{n\_estimators} parameter specifies the number of trees in the forest, where more trees generally improve performance but increase computational cost. The \texttt{max\_depth} parameter controls the maximum depth of each tree, preventing overfitting by limiting how deep the trees can grow. The \texttt{max\_features} parameter determines the number of features to consider when looking for the best split, balancing between the model's accuracy and computational efficiency.

%% RANDOM FOREST HYPERPARAMS %%
\begin{table}[ht!]
\centering
\begin{tabular}{c|c|c|c|c}
\hline
\textbf{Model} & 
\# features ($p$) & \texttt{n\_estimators} & \texttt{max\_depth} & \texttt{max\_features} \\
\hline
\multirow{3}{*}{Random Forest} 
 & $p=2$ & 72 & 40 & sqrt \\
 & $p=3$ & 197 & 96 & sqrt \\
 & $p=10$ & 103 & 13 & sqrt \\
\hline
\end{tabular}

\caption{Optimal values for the hyperparameters of the Random Forest classifier found by the randomized search}
\label{tab:rf_hyperparams}

\end{table}

In order to ensure the reproducibility of the results, all of these classical experiments have been conducted using the \texttt{scikit-learn} Python library.

\end{document}